# Understanding the Relationship Between Firms' AI Technology Innovation and Consumer Complaints

**Yongchao Martin Ma**, *Senior Member, IEEE**
School of Management
Huazhong University of Science & Technology, PR China
Department of Marketing
City University of Hong Kong, Hong Kong

**Zhongzhun Deng**[†]
Business School
Sichuan University, Sichuan, PR China

March 5, 2026

## Abstract

In the artificial intelligence (AI) age, firms increasingly invest in AI technology innovation to secure competitive advantages. However, the relationship between firms' AI technology innovation and consumer complaints remains insufficiently explored. Drawing on Protection Motivation Theory (PMT), this paper investigates how firms' AI technology innovation influences consumer complaints. Employing a multi-method approach, Study 1 analyzes panel data from S&P 500 firms ($N$ = 2,758 firm-year observations), Study 2 examines user-generated Reddit data ($N$ = 2,033,814 submissions and comments), and Study 3 involves two controlled experiments ($N$ = 410 and $N$ = 500). The results reveal that firms' AI technology innovation significantly increases consumers' threat-related emotions, heightening their complaints. Furthermore, compared to AI process innovation, AI product innovation leads to higher consumer complaints. This paper advances the understanding of consumers' psychological responses to firms' AI innovation and provides practical implications for managing consumer complaints effectively.

***Keywords*** Artificial intelligence · Product and process innovations · Consumer complaints · Protection motivation theory · Threat emotions

*The authors acknowledge financial support from the National Natural Science Foundation of China (project 72202149) and the China Postdoctoral Science Foundation – Hubei Joint Support Program under Grant Number 2025T100HB for the research, authorship, and/or publication of this article. The computation is completed in the HPC Platform of Huazhong University of Science and Technology.

[†]Corresponding author. Email: `danieldeng@scu.edu.cn`

# Understanding the Relationship Between Firms' AI Technology Innovation and Consumer Complaints

**Abstract**

In the artificial intelligence (AI) age, firms increasingly invest in AI technology innovation to secure competitive advantages. However, the relationship between firms' AI technology innovation and consumer complaints remains insufficiently explored. Drawing on Protection Motivation Theory (PMT), this paper investigates how firms' AI technology innovation influences consumer complaints. Employing a multimethod approach, Study 1 analyzes panel data from S&P 500 firms (N = 2,758 firm-year observations), Study 2 examines user-generated Reddit data (N = 2,033,814 submissions and comments), and Study 3 involves two controlled experiments (N = 410 and N = 500). The results reveal that firms' AI technology innovation significantly increases consumers' threat-related emotions, heightening their complaints. Furthermore, compared to AI process innovation, AI product innovation leads to higher consumer complaints. This paper advances the understanding of consumers' psychological responses to firms' AI innovation and provides practical implications for managing consumer complaints effectively.

**Keywords:** Artificial intelligence, product and process innovations, consumer complaints, protection motivation theory, threat emotions

## 1. Introduction

Artificial intelligence (AI) has fundamentally transformed digital business, revolutionizing personalized marketing and consumer service dynamics (Ameen et al., 2022b; Wirtz & Pitardi, 2023). Recently, McKinsey forecasts that AI can enhance firms' marketing productivity by up to 15% of total marketing expenditures, amounting to approximately $463 billion annually.[1]

AI technologies improve production and operational efficiencies and significantly enhance consumer experiences (Grewal et al., 2024; Hermann & Puntoni, 2024). Companies increasingly invest in AI technologies by embedding AI features into their products and services or associating their offerings explicitly with AI labels. For example, Samsung recently introduced the French Door Refrigerator, which employs AI technology to enhance energy efficiency and simplify everyday usage.[2]

The impact of AI as a transformative general-purpose technology on firm performance has emerged as a critical area of inquiry (Alderucci et al., 2019; Cockburn et al., 2019; Yang, 2022). Firms' investment in AI technologies (hereafter "AI innovation") has endowed AI with remarkable capabilities across diverse domains by leveraging extensive training data, facilitating AI's increasing integration into consumers' daily lives (Ameen et al., 2022b; Peres et al., 2023). However, AI poses significant challenges to fundamental public rights, including bias prevention, personal freedom, privacy, and employment (Gerards, 2019), potentially triggering negative consumer responses.

Consumer complaints, defined as behavioral "expressions of dissatisfaction" (Kowalski 1996, p. 180), negatively impact firms' marketing performance and image (Gregoire et al., 2024). Despite this importance, existing research inadequately addresses the relationship between firms' AI innovation and consumer complaints, particularly regarding effective mitigation strategies.

---

[1] https://www.mckinsey.com/capabilities/growth-marketing-andsales/our-insights/.how-generative-ai-can-boost-consumer-marketing
[2] https://finance.sina.com.cn/tech/digi/2024-08-28/doc-incmesvh5038009.shtml

Additionally, few studies differentiate consumer responses based on whether AI technology enhances intangible processes (e.g., backend operations, hereafter "AI process innovation") or tangible products (e.g., consumer-facing applications, hereafter "AI product innovation") (Frank, 2024; Frank et al., 2021). In response to these gaps, this study investigates three research questions (RQs):

*RQ1: Does firms' AI innovation increase consumer complaints about firms?*

*RQ2: What is the underlying mechanism behind this relationship?*

*RQ3: Does the type of AI innovation type (process vs. product) influence consumer complaints?*

We employ the Protection Motivation Theory (PMT) to examine how AI innovation influences consumer complaints. Specifically, we propose that firms' AI innovation heightens consumers' negative emotional responses concerning threats posed by AI technology (hereafter, "threat emotion"), which in turn increases consumer complaints about firms. Additionally, firms invest in AI technologies not only to enhance productivity in production and operations but also to integrate AI features into their products and services. However, given that product technology is more visible to consumers than process technology (Un & Asakawa, 2015), we further hypothesize that AI product innovation amplifies consumers' threat emotions, intensifying the effect on consumer complaints.

Our research consists of three complementary studies: In Study 1, we analyze panel data from S&P 500 companies to explore how AI innovation affects consumer complaints, demonstrating that this effect is more pronounced for AI product innovation than for AI process innovation. Study 2 utilizes RoBERTa-based sentiment analysis on Reddit data to examine how ChatGPT, a leading recent AI innovation, affects consumers' feelings of threat and their subsequent complaints. Finally, Study 3 consists of two controlled experiments designed to validate the mediating role of threat emotions and replicate the key findings from Studies 1 and 2.

The remainder of this paper is structured as follows: First, we present the theoretical foundation and conceptual framework. Next, we detail the methodologies and findings of our three empirical studies. We conclude by discussing theoretical implications and suggesting avenues for future research.

## 2. Literature Review

### 2.1 Firms' AI Innovation

Existing literature emphasizes that AI technologies generate bidirectional effects, entailing both benefits and challenges for firms (André et al., 2018; Grewal et al., 2021). Prior research identifies several domains where AI demonstrates significant positive influences. At the individual level, AI enhances personal productivity and enriches daily experiences (Liang et al., 2021; Thamik & Wu, 2022). At the firm level, AI significantly improves operational efficiency and optimizes business processes (Latinovic & Chatterjee, 2022; Li & Xu, 2022; Siemon, 2022; Yen & Chiang, 2020).

The past decade has witnessed unprecedented advancements in AI technologies, notably in machine learning, robotics, and neural networks (Trajtenberg, 2018). Consequently, understanding the impact of rapid AI technology development on firm performance has become increasingly critical (Alderucci et al., 2019; Cockburn et al., 2019; Yang, 2022). AI enhances operational efficiency and improves production processes (Gans, 2024; Prahl & Goh, 2021). Thus, firms seeking a competitive advantage increasingly prioritize AI development by intensifying investments and applying for relevant patents (Alderucci et al., 2019; Yang, 2022).

Meanwhile, firms' investments in AI technology have created complex dynamics in consumer responses (Chandra et al., 2022; Darveau & Cheikh-Ammar, 2021). On the one hand, recent advances in generative AI, notably exemplified by ChatGPT, have demonstrated unprecedented capabilities in natural language processing, computer programming, interactive dialogue, and complex problem-solving (Chen et al., 2023). These AI technologies, trained on extensive datasets (Peres et al., 2023), facilitate human-like communication through emotional intelligence and interactive functionalities (Huang & Rust, 2024). Furthermore, fine-tun foundational models with proprietary firm data enable businesses to capture brand-specific characteristics and derive deeper consumer insights (Huang & Rust, 2024).

On the other hand, firms also face significant challenges due to consumers' negative responses to AI technological advancements. For instance, firms must balance safeguarding consumer privacy and maintaining competitive advantages (Cillo & Rubera, 2024). While advanced AI technologies offer substantial opportunities, they raise important psychological concerns simultaneously. Marriott and Pitardi (2024) demonstrate that user loneliness and fear of judgment, combined with perceived AI sentience and associated well-being benefits, increase tendencies toward app addiction. Pinto et al. (2023), examining student acceptance of large language models (LLMs), find widespread adoption accompanied by relatively low technology anxiety, primarily focused on concerns about AI-related job displacement. Moreover, AI technologies may evoke various negative consumer reactions. Ma et al. (2024) provided a systematic framework identifying 12 distinct negative impacts of AI (see Figure 2 in Ma et al., 2024).

Given the potential detrimental effects of negative consumer responses, such as increased complaints that harm brand evaluations and sales performance (e.g., Babić Rosario et al. 2016), it is crucial to understand how firms' AI innovation strategies influence these responses. Therefore, we specifically investigate how firms' AI innovation affects consumer complaints.

## 2.2 Consumer Complaints

A consumer complaint refers to a behavioral "expression of dissatisfaction … for venting emotions or achieving intrapsychic goals, interpersonal goals, or both" (e.g., emotional relief, compensation, or social positioning) (Kowalski 1996, p. 180). Consumers articulate their dissatisfaction through various channels (Ward and Ostrom 2006), among which publicly visible complaints that reach numerous observers can negatively impact brand evaluations and sales (e.g., Babić Rosario et al. 2016).

Previous research has mainly focused on response strategies to mitigate the negative impacts of consumer complaints. For example, prior studies suggest that responding to complaints, regardless of the specific strategy employed, typically results in more favorable outcomes than remaining silent, as observers interpret brand responses as indicators of respect, professionalism, and willingness to engage (Weitzl & Hutzinger, 2019). Furthermore, accommodative response strategies have been identified as particularly effective, fostering positive consumer outcomes such as greater satisfaction, improved brand evaluations, and increased purchase intentions (Chang et al. 2015; Xia 2013).

However, limited prior research has examined consumer complaints about firms' AI innovation. Moreover, existing studies have not specifically investigated how different types of AI innovation (i.e., AI process innovation versus AI product innovation) interact to influence consumer complaints. This research gap motivates our current investigation into the effects of AI innovation types on consumer complaint behaviors.

**2.3 Protection Motivation Theory**

Protection Motivation Theory (PMT), introduced by Rogers (1975), provides a robust theoretical framework for understanding how individuals evaluate and respond to perceived threats by adopting protective measures. According to PMT, protective behaviors are influenced primarily by two key components: threat appraisal (which encompasses perceived vulnerability and perceived severity) and coping appraisal (including response efficacy, self-efficacy, and response costs).

PMT has gained prominence in information systems and marketing research as an effective framework for examining protective behavioral responses (Cram et al., 2019; Moody et al., 2018). Its applicability extends across contexts where understanding and influencing protective behaviors are critical (Shore et al., 2022), including recent applications related to AI technology and consumer attitudes (Al-Sharafi et al., 2023; Arpaci, 2024).

In the context of our research, firms increasingly invest in AI technologies to achieve competitive advantages (Sáez-Ortuño et al., 2024). However, the growing investment and proliferation of AI technologies can heighten consumer perceptions of threat, potentially triggering negative emotions and increasing consumer complaints toward firms.

PMT provides an appropriate theoretical lens to explore this relationship, emphasizing consumers' vulnerability and threat severity appraisals. Specifically, threat severity appraisal helps analyze how intensely consumers perceive the potential negative consequences of firms' AI innovation. Concurrently, vulnerability appraisal allows examination of how consumer perceptions differ based on the type of AI innovation (e.g., process vs. product technology). Thus, applying PMT illuminates the mechanisms underlying consumers' protective responses, such as increased complaints, in reaction to firms' AI technology innovation.

## 3. Conceptual Framework and Hypothesis Development

### 3.1 Firm Innovation, Threat Appraisal, and Complaints

Threat appraisal, a key component of protective responses, comprises two fundamental elements: perceived severity and vulnerability. Perceived severity refers to individuals' assessments regarding the seriousness of potential consequences arising from a threat, whereas perceived vulnerability represents individuals' judgments about their susceptibility to encountering the threat (Floyd et al., 2000).

Recent advancements in sophisticated AI model training and extensive task automation capabilities have expanded AI applications into increasingly complex domains. Examples include ChatGPT (a large language model) utilized for writing tasks, Stable Diffusion (a deep-learning-based algorithm) applied in image generation, and Sora (a generative AI model), which is capable of producing videos from textual descriptions. Such developments may heighten consumers' threat appraisals by increasing perceptions of the severity and vulnerability associated with AI technologies, potentially leading to heightened consumer complaints.

Firms seeking a competitive advantage increasingly prioritize AI development through intensified investments and patent applications (Alderucci et al., 2019; Yang, 2022), substantially accelerating AI technology advancements. Investments in AI technologies by these firms influence consumers' threat appraisal processes in various ways.

First, firms' AI innovation increases consumers' perceptions of threat severity. For instance, generative AI requires extensive collection and utilization of personal information, significantly intensifying data collection practices (Puntoni et al., 2021). Heightened privacy threats are commonly associated with technologies demanding extensive personal information disclosure (Youn, 2009).

Second, AI innovation elevates consumers' perceptions of vulnerability. New AI technologies are increasingly integrated into daily applications (Peres et al., 2023), making it increasingly challenging for consumers to protect their data privacy, raising their perceived vulnerability. Additionally, firms' AI innovation intensifies consumer concerns about job displacement (Granulo et al., 2019) and algorithmic bias (Arnold et al., 2024), further amplifying overall threat perceptions.

Consequently, consumers' perceptions of threat severity and vulnerability intensify as firms escalate their investments in AI technology. Perceived vulnerability involves individuals assessing the likelihood of encountering a threat (Zhang et al., 2018), causing them to perceive themselves as more susceptible. As consumers perceive greater threats, they are likely to experience stronger negative emotions. Thus, we propose the following hypothesis:

*$H_1$: Firms' AI technology innovation increases consumers' threat emotions.*

Consumer complaints involve substantial psychological resources related to negative emotions (e.g., anger) and cognitions (e.g., betrayal)(Gregoire et al., 2024). This study addresses consumers' negative emotional responses resulting from firms' innovation in AI technology. Firms' increased investments in AI accelerate technological advancements and intensify consumers' negative threat-related emotions, such as fear, anger, disgust, and disapproval. These threat-induced negative emotions mobilize substantial psychological resources to increasing consumers' likelihood of complaining (Gregoire et al., 2024). Therefore, we propose the following hypothesis:

*H₂: (a) Firms' AI technology innovation increases consumer complaints toward firms, and (b) heightened consumer threat emotions mediate this relationship.*

**3.2 Moderating Effects of Innovation Type**

We propose that consumers' threat emotions mediate the relationship between AI innovation and increased consumer complaints. Factors influencing consumers' threat perceptions are likely to moderate these effects. Based on this logic, we examine a critical moderating factor: the type of AI innovation.

AI technologies were initially adopted to empower firms' production and operational processes, which tend to be intangible and less visible to consumers (Frank et al., 2021). However, recent advancements have enabled AI technologies to empower tangible consumer products (André et al., 2018). Several generative AI tools, such as ChatGPT and Stability AI, exemplify this shift toward visible, consumer-facing products (Stokel-Walker & Van Noorden, 2023).

Drawing on the distinction between product and process innovations (Cohen & Klepper, 1996a, 1996b; Scherer, 1984; Scherer, 1982), we distinguish between two types of AI innovation. AI process innovation involves implementing novel manufacturing or operational methods for existing products, characterized by an internal, efficiency-driven focus (Terjesen & Patel, 2015). Such technologies create value by improving internal production processes (Kammerer, 2009) or delivery methods (Crossan & Apaydin, 2010; Damanpour, 2010). In contrast, AI product innovation focuses on developing new consumer-facing products driven by external market demands. Due to their inherent consumer visibility, AI product innovation offers direct observability to external stakeholders (e.g., consumers) (Un & Asakawa, 2015).

Considering the visibility differential between product and process innovations, these innovation types may differentially influence the visibility of AI technology and, thus, consumers' perceived threats. Specifically, because AI product innovation is more visible, they amplify consumers' threat perceptions and complaints more significantly than AI process innovation. Therefore, we propose that AI product innovation intensifies the effect of AI innovation on consumers' threat emotions and subsequent complaints:

*$H_3$: AI product (vs. process) innovation leads to stronger consumer threat emotions, enhancing the relationship between AI technology innovation and consumer complaints.*

Figure 1 illustrates our conceptual model.

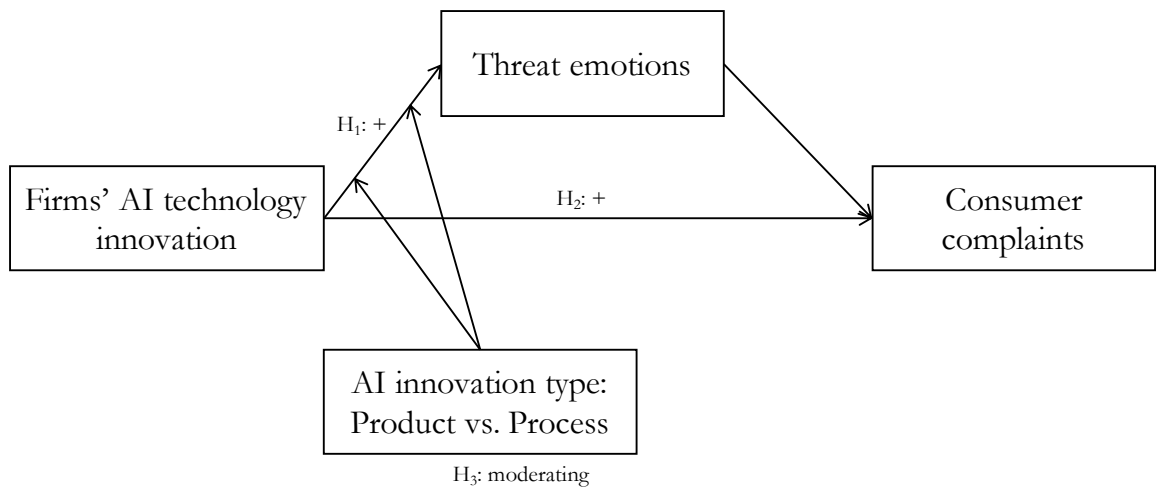

**Figure 1**. Conceptual Model and Study Design

## 4. Overview of Empirical Studies

Our research employs a comprehensive multimethod design to test our proposed hypotheses rigorously. A summary of these studies is presented in Table 1.

In Study 1, we analyze panel data from S&P 500 firms to investigate the direct relationship between firms' AI innovation and consumer complaints and the moderating effect of innovation type. Study 2 uses Reddit data to explore how consumer threat emotions influence consumer complaints. This study employs advanced analytical methods, including RoBERTa-based sentiment analysis and text mining techniques, to capture consumer sentiment and complaint behaviors effectively. In Study 3, we conduct controlled experiments to validate the mediating role of threat emotions, replicating and reinforcing the findings of Studies 1 and 2.

**Table 1.** Summary of Studies

| Study | Design | Samples, Source | DV(s) | Hypotheses |
|---|---|---|---|---|
| 1 | Secondary Data study using firm data | N = 379, Databases | Complaints | $H_{2a}$, $H_3$ |
| 2 | Secondary Data study using social media data | N = 2,033,814, Reddit data | Complaints | $H_1$, $H_{2b}$ |
| 3a | 2 (AI innovation: Emphasizing AI innovation vs. control group) × 2 (Innovation type: process vs. product) | N = 355, Prolific | Complaints | $H_1$, $H_{2a}$, $H_{2b}$, $H_3$ |
| 3b | 2 (AI innovation: Emphasizing AI innovation vs. control group) × 2 (Innovation type: process vs. product) × 2 (Threat manipulation: manipulation vs. control group) | N = 393, Prolific | Complaints | $H_1$, $H_{2a}$, $H_{2b}$, $H_3$ |

## 4.1 Study 1: Analysis of the Relationship Between Firms' AI Innovation and Consumer Complaints Using S&P 500 Data

### 4.1.1 Data Collection and Sample

Our analysis focuses on S&P 500 listed firms from 2009 to 2021. We applied several exclusion criteria to ensure the validity and consistency of our sample. First, we excluded firms from the financial sector (Global Industry Classification Standard codes 4010, 4020, and 4030) due to their distinctive innovation patterns (Jiang et al., 2023). Second, to maintain homogeneity in financial and economic development, legal frameworks, and public infrastructure, we restricted our sample exclusively to U.S.-based firms (Hilary & Hui, 2009). Finally, we removed observations lacking complete regression variables.

The final dataset integrates information from three primary sources: (1) the ASSET4 database for consumer complaints; (2) the United States Patent and Trademark Office (USPTO)'s PatentsView database for innovation-related metrics; and (3) the Orbis and ASSET4 databases for firms' basic financial and ESG information. The resulting sample consists of 2,758 firm-year observations across 379 distinct S&P 500 firms from 2009 to 2021.

**4.1.2 Measurement**

**Consumer Complaints.** We measured consumer complaints (*Complaints*) using data from the Thomson Reuters ASSET4 database (Mitra et al., 2021), which ensures data quality through rigorous validation processes (LSEG Data & Analytics, 2023). Specifically, consumer complaints were measured based on their annual frequency for each firm (Ioannou et al., 2023). Higher values indicate a greater number of complaints.

**AI Innovation.** Patents reflect firms' strategic investments in technology (Ma et al., 2023; Yang, 2022), providing a reliable measure of AI-related investments and developments. Following Yang (2022), we measured firms' AI innovation (*AI innovation*) using the annual count of AI-related patents granted to each firm. These patents were identified using the USPTO classification system, specifically under U.S. Patent Classification System (USPC) codes 706 (Artificial Intelligence) and 382 (Image Analysis). Consistent with Jiang et al. (2023), firms without successful USPTO patent applications were assigned a zero value.

**Innovation Type.** Since the USPTO patent dataset does not explicitly distinguish between product and process innovations, we adopted the patent classification methodology developed by Ganglmair et al. (2022) and Bena (2023). This method classifies independent patent claims into either process- or product-related categories based on textual analysis of claim descriptions. Specifically, patent claims containing introductory phrases such as "apparatus," "device," "machine," "computer," "circuit," or "semiconductor" were classified as process-related. In contrast, claims containing terms such as "method," "process," "approach," "manner," "practice," "technique," or "treatment" were classified as product-related. We quantified *Innovation type* as the ratio of product claims to total claims in patents granted to firm *i* in year *t*.

**Control Variable.** We included several firm-level control variables to account for financial and ESG performance factors that may influence consumer complaints. Additionally, we incorporated industry and year fixed effects to address unobservable heterogeneity.

We winsorized all continuous variables at the 1st and 99th percentiles to mitigate the impact of outliers. Variance inflation factor (VIF) analysis yielded a value of 1.41, indicating no significant multicollinearity concerns. Table 2 summarizes all variables, while Table 3 presents pairwise correlations.

**TABLE 2.** Variable Summary

| Variable | Definition | Data source |
|---|---|---|
| **Dependent variables** | | |
| Complaints | Number of consumer controversies in the focal year | ASSET4 |
| **Independent variables** | | |
| AI innovation | Number of AI patents in the focal year | UPSTO, KPSS |
| **Moderators** | | |
| Innovation type | The ratio of product claims to all patent claims in the focal year | UPSTO, KPSS |
| **Control variables** | | |
| **Firm level** | | |
| Roa | Natural logarithm of firm's return on assets ratio | Orbis |
| Size | Natural logarithm of the firm's total employees | Orbis |
| Age | Number of years since IPO | Orbis |
| Other patents | Natural logarithm of total patents | UPSTO, KPSS |
| Market cap | Natural logarithm of total market capitalization | Orbis |
| Growth | The growth rate of total income | Orbis |
| ISO | A dummy variable that equals one if the firm attained ISO14001 certification and zero otherwise | ASSET4 |
| CSR reporting | A dummy variable that equals one if the firm reports CSR performance and zero otherwise | ASSET4 |

**Table 3**. Descriptive Statistics and Correlations

| | 1 | 2 | 3 | 4 | 5 | 6 | 7 | 8 | 9 | 10 | 11 |
|---|---|---|---|---|---|---|---|---|---|---|---|
| 1. Complaints | 1 | | | | | | | | | | |
| 2. AI innovation | 0.21*** | 1 | | | | | | | | | |
| 3. Product type | 0.02* | 0.11*** | 1 | | | | | | | | |
| 4. Roa | 0.04*** | 0.06*** | 0.09*** | 1 | | | | | | | |
| 5. Size | 0.10*** | 0.27*** | -0.0100 | 0.15*** | 1 | | | | | | |
| 6. Age | -0.02 | 0.13*** | -0.09** | 0.15*** | 0.28*** | 1 | | | | | |
| 7. Other patents | 0.10*** | 0.62*** | 0.09*** | 0.09*** | 0.30*** | 0.28*** | 1 | | | | |
| 8. Market cap | 0.17*** | 0.38*** | -0.0100 | 0.14*** | 0.58*** | 0.27*** | 0.39*** | 1 | | | |
| 9. Growth | 0.02 | -0.04** | 0.11*** | 0.08*** | -0.22*** | -0.22*** | -0.03** | -0.12*** | 1 | | |
| 10. ISO | 0 | 0.17*** | 0.0500 | 0.09*** | 0.15*** | 0.20*** | 0.37*** | 0.15*** | -0.09*** | 1 | |
| 11. CSR reporting | 0.05*** | 0.14*** | -0.06* | -0.0100 | 0.31*** | 0.27*** | 0.19*** | 0.41*** | -0.22*** | 0.29*** | 1 |
| Mean | 0.12 | 0.22 | 0.51 | 1.58 | 9.73 | 2.95 | 1.68 | 9.84 | 0 | 0.33 | 0.66 |
| SD | 1.2 | 0.68 | 0.29 | 1.23 | 1.53 | 0.95 | 2.19 | 1.15 | 0.01 | 0.47 | 0.47 |
| Min | 0 | 0 | 0 | -3.01 | 5.23 | 0 | 0 | 7.19 | -0.03 | 0 | 0 |
| Max | 54 | 6.02 | 1 | 3.36 | 12.9 | 4.54 | 7.46 | 12.69 | 0.05 | 1 | 1 |

*Notes*: Correlations are bold at the $p < .10$ level.

### 4.1.3 Analysis Method

Given the distinctive characteristics of our dependent variable, *Complaints*, a continuous variable exhibiting numerous zero values, a specialized analytical approach was required.

We employed a generalized linear model (GLM) using a Poisson distribution with a log-link function, incorporating robust standard errors to address potential heteroscedasticity and non-normality (Kraft et al., 2024).

Additionally, we performed supplementary regression analyses as robustness checks to validate the consistency of our findings. All independent and control variables were lagged by one year to account for temporal delays between firms' strategic decisions and observed consumer complaint outcomes. The analyses were conducted using Stata 18.

### 4.1.4 Results

Our analysis revealed several significant findings, as summarized in Table 4. Models 1 and 2 report the main effects of *AI innovation*. Model 1 includes only the focal independent variable (*AI innovation*), while Model 2 incorporates fixed effects and control variables. In Model 2, the coefficient for *AI innovation* was significantly positive ($\beta = 0.548$, $p = .001$), supporting $H_{2a}$.

Models 3 and 4 examine the moderating effect of *Innovation type* (product vs. process). We mean-centered the *Innovation type* before creating interaction terms to improve interpretability (Aiken & West, 1991). Results from Model 4 indicate a significant positive interaction effect between *AI innovation* and *Innovation type* ($\beta = 0.924$, $p = .009$).

Simple slope analysis, using ±1 standard deviation from the mean to define high and low levels of *Innovation type* (L. Wang et al., 2023), demonstrates that the relationship between AI innovation and complaints is significantly stronger at higher levels of product innovation (see Figure 2). These findings support $H_3$.

**Table 4**. Results of GLM Regression

| DV | | | Complaints | |
|---|---|---|---|---|
| | **Model 1** | **Model 2** | **Model 3** | **Model 4** |
| *ROA* | | 0.108 | 0.042 | -0.017 |
| | | (0.174) | (0.169) | (0.164) |
| | | [0.534] | [0.802] | [0.916] |
| *Size* | | 0.004 | 0.080 | 0.121 |
| | | (0.126) | (0.115) | (0.115) |
| | | [0.974] | [0.486] | [0.292] |
| *Age* | | 0.028 | 0.046 | 0.047 |
| | | (0.156) | (0.157) | (0.151) |
| | | [0.855] | [0.771] | [0.757] |
| *Other patents* | | -0.049 | -0.049 | -0.069 |
| | | (0.117) | (0.118) | (0.111) |
| | | [0.673] | [0.679] | [0.535] |
| *Market cap* | | 0.742*** | 0.760*** | 0.758*** |
| | | (0.148) | (0.146) | (0.146) |
| | | [0.000] | [0.000] | [0.000] |
| *Growth* | | -12.453 | -14.466 | -12.320 |
| | | (14.480) | (14.386) | (14.042) |
| | | [0.390] | [0.315] | [0.380] |
| *ISO* | | 0.341 | 0.414 | 0.403 |
| | | (0.348) | (0.366) | (0.368) |
| | | [0.328] | [0.258] | [0.274] |
| *CSR reporting* | | 0.120 | 0.003 | -0.011 |
| | | (0.483) | (0.472) | (0.472) |
| | | [0.804] | [0.995] | [0.981] |
| *AI innovation* | 0.824*** | 0.548*** | 0.448*** | 0.471*** |
| | (0.070) | (0.160) | (0.165) | (0.157) |
| | [0.000] | [0.001] | [0.007] | [0.003] |
| *Innovation type* | | | -1.187* | 0.863 |
| | | | (0.635) | (0.981) |
| | | | [0.061] | [0.379] |
| *AI innovation × Innovation type* | | | | 0.924*** |
| | | | | (0.353) |
| | | | | [0.009] |
| _cons | -2.753*** | -29.080*** | -29.495*** | -30.828*** |
| | (0.140) | (1.622) | (1.881) | (1.621) |
| | [0.000] | [0.000] | [0.000] | [0.000] |
| Year FE | No | Yes | Yes | Yes |
| Industry FE | No | Yes | Yes | Yes |
| Observations | 2,758 | 2,758 | 2,758 | 2,758 |
| Log-likelihood | -1966.208 | -785.368 | -779.672 | -770.035 |

*Notes*: Robust standard errors clustered at the firm level are reported in parentheses, and exact p-values are reported in square brackets.
Standard errors in parentheses; * $p < 0.1$, ** $p < 0.05$, *** $p < 0.01$.

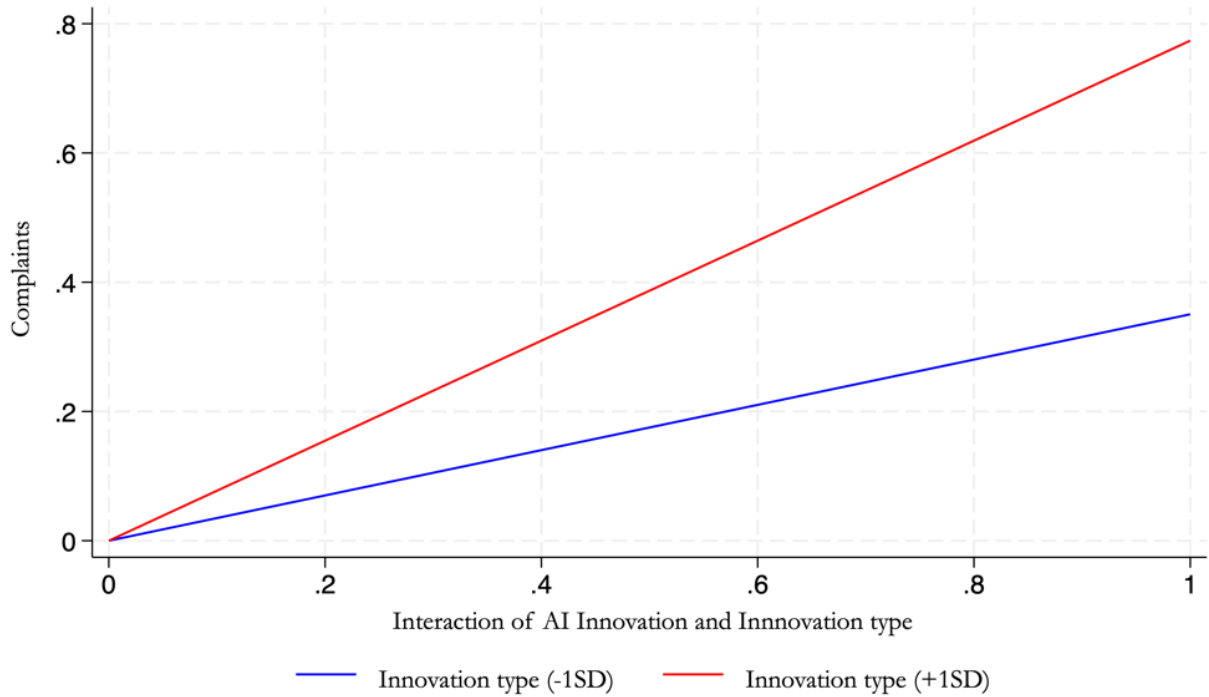

**Figure 2**. The Interaction Between AI Innovation and Product Type (Study 1)

**4.1.5 Robustness Analyses**

To further validate the robustness of our findings, we expanded our analysis to encompass all S&P 500 firms available in the KPSS database, assigning zero patents to firms without patent applications (Kraft et al., 2024). This expanded analysis confirmed that our results are not subject to sample selection bias. Additionally, we conducted several supplementary robustness tests to address potential measurement errors and omitted variable bias.

First, we tested alternative measurements for our dependent and independent variables. Specifically, we used the raw count of AI patents based on the International Patent Classification (IPC) codes as an alternative measure of AI innovation (Yang, 2022). These alternative measurements consistently supported our hypotheses. Detailed results are presented in Web Appendix B.

Next, we employed the Robustness of Inference to Replacement (RIR) method to assess the potential influence of omitted variable bias. This analysis indicated that invalidating our findings would require replacing 60.41% of cases with null effects, demonstrating strong robustness against potential confounding variables.

Overall, these comprehensive robustness analyses substantiate our main findings on the relationship between firms' AI innovation and consumer complaints and the moderating role of AI innovation type. Specifically, while firms' AI innovation increases consumer complaints toward firms, this relationship is amplified for AI product innovation. These results have important implications for downstream marketing outcomes. The following section further investigates evidence supporting the proposed underlying mechanism.

**4.2 Study 2: Analysis of the Relationship Between Threat Emotion and Complaints Using Reddit Data**

Study 2 utilizes real-world data to examine the relationship between consumers' threat emotions and their complaints regarding AI technology. This analysis provides further evidence supporting the mediating effect of threat emotions proposed in our conceptual model. We focus specifically on OpenAI's generative AI chatbot, ChatGPT, which was launched in 2022. ChatGPT provides an ideal research setting due to OpenAI's sequential releases of increasingly sophisticated model versions, facilitating the observation of evolving consumer reactions[3].

Following established methodological approaches, we analyzed social media data collected throughout 2023 to investigate how consumer threat emotions relate to consumer complaints (Vaid et al., 2023). We predict a positive relationship, such that heightened threat emotions are associated with increased complaints.

---

[3] https://en.wikipedia.org/wiki/ChatGPT

### 4.2.1 Data Collection

Reddit serves as a valuable source of user-generated content, ranking as the fifth most visited website in the United States with over 430 million monthly active users (as of September 2020). The platform's structure into topic-specific "subreddits" creates naturally segmented communities, enabling focused analyses of user discussions. Reddit's diverse user base and segmented organization allow us to capture various consumer perspectives on AI technologies.

We selected the r/chatgpt subreddit for our analysis based on two primary criteria: its direct relevance to discussions about ChatGPT and its high user engagement metrics, with over 8 million members (ranking in the top 1% of Reddit communities).[4]

The final dataset consists of 176,167 submissions and 1,857,647 comments collected throughout 2023, offering a comprehensive sample of user discussions within this ChatGPT-focused community.

### 4.2.2 Measurement

*Complaints.* We utilized a semi-supervised guided Latent Dirichlet Allocation (LDA) model to classify submissions as complaints or not based on the dominant topic identified in users' descriptions of their experiences (see Table 5; Astvansh et al., 2024).

**Table 5.** List of words

| Topic | List of words |
|---|---|
| Complaints | Disappoint, disappointed, tired, unhappy, unacceptable, unbelievable, bother, bothered, dissatisfied, trouble, inconvenience, inconvenient, annoyed, annoying, fault, irritate, irritated, difficulty, difficult, unpleasant, ruined, upset, terrible, awful, horrible, careless, negligent, negligence, distressing, distressed, stress, mad, scared, scary, complaint, complaint, sue, charge, furious, cost, money, expense, expensive, purchase, buy, pay, annoy, worried, worry, late, satisfaction, satisfied, dissatisfied, trust, displeased, unfortunate, unsatisfied, adverse, unfavorable, distrust, wary, doubt, angry. |

*Source:* Astvansh et al. (2024).

*Threat emotions.* To measure discrete emotional responses within Reddit submissions, we analyzed sentiment using the RoBERTa-base model, fine-tuned on the go_emotions dataset. Recent advancements in automated text mining have significantly improved sentiment analysis techniques (Ma et al., 2024), with BERT-based models, including RoBERTa, demonstrating superior capabilities in contextual understanding (Min et al., 2021). Specifically, we employed the RoBERTa architecture explicitly trained for multilabel emotional classification across 28 dimensions (Hartmann et al., 2023; Liu et al., 2019). Detailed sentiment classifications and their mean values are provided in Web Appendix B.

---
[4] https://www.reddit.com/r/ChatGPT/?rdt=48410

Our measurement of threat emotions aligns with PMT's conceptualization of fear appeals (Rogers, 1975), which posits that fear-inducing messages alter individuals' perceptions of self-efficacy, response efficacy, threat severity, and susceptibility. Such perceptions subsequently influence behavioral attitudes through heightened perceived threats and fear levels. Consistent with prior literature (Moody et al., 2018), we operationalized *Threat emotion* by aggregating the emotional scores for *fear*, *anger*, *disgust*, and *disapproval*.

*Control Variables.* Following prior research suggesting content-related factors significantly influence user-generated content (Gu et al., 2023; Zhang & Luo, 2022), we included several controls in our analysis: word count (length of the submission), image (a binary indicator coded as one if the submission included a photo, otherwise 0), score (the submission's engagement score), and comment count (the number of comments per submission). Additionally, we incorporated month-fixed effects to control for temporal trends.

**4.2.3 Regression Analysis**

Table 6 summarizes our comprehensive regression analysis. In Model 1, we observe a positive and statistically significant relationship between *Threat emotion* and *Complaints* ($\beta = 0.231$, $p < .001$), providing strong support for $H_1$. This significant relationship remains robust after including submission-level control variables and month-fixed effects.

To further validate these findings, we separately analyzed discrete threat-related emotions—*fear*, *anger*, *disgust*, and *disapproval*—in Models 2–4. Results from these analyses remained consistent with our primary assumptions, providing further empirical evidence supporting our theoretical framework, which links consumer threat emotions to increased complaints.

Table 6. Regression Results on Technology Anxiety

| DV | Complaints | | | | |
|---|---|---|---|---|---|
| | Model 1 | Model 2 | Model 3 | Model 4 | Model 5 |
| Threat emotions | 1.501*** | | | | |
| | (0.096) | | | | |
| | (0.000) | | | | |
| Fear | | 0.986*** | | | |
| | | (0.140) | | | |
| | | (0.000) | | | |
| Anger | | | 0.579*** | | |
| | | | (0.112) | | |
| | | | (0.000) | | |
| Disgust | | | | 0.237*** | |
| | | | | (0.010) | |
| | | | | (0.000) | |
| Disapproval | | | | | 0.158*** |
| | | | | | (0.010) |
| | | | | | (0.000) |
| _cons | -2.357*** | -0.994*** | -1.528*** | -2.122*** | -1.795*** |
| | (0.026) | (0.026) | (0.028) | (0.019) | (0.019) |
| | (0.000) | (0.000) | (0.000) | (0.000) | (0.000) |
| Month FE | Yes | Yes | Yes | Yes | Yes |
| Baseline control | Yes | Yes | Yes | Yes | Yes |
| Observations | 2,033,814 | 2,033,814 | 2,033,814 | 2,033,814 | 2,033,814 |
| Log-likelihood | 195690.966 | 200693.879 | 204321.775 | 195757.988 | 199505.795 |
| AIC | -391355.931 | -401361.758 | -408615.550 | -391489.975 | -398985.590 |
| BIC | -391224.927 | -401230.728 | -408474.468 | -391358.970 | -398854.561 |

*Notes*: Robust standard errors clustered at the submission level are reported in parentheses, and exact p-values are reported in square brackets. In brief, we did not report the coefficients of control variables.
Standard errors in parentheses; * $p < 0.1$, ** $p < 0.05$, *** $p < 0.01$.

### 4.2.4 Discussion

Study 2 provides empirical support for $H_1$ by establishing a clear positive relationship between consumers' threat emotions and complaints. These findings reinforce our theoretical model and provide preliminary evidence supporting the proposed mediating role of threat emotions. However, further investigation is necessary to robustly confirm this mediation effect ($H_{2b}$). Thus, in Study 3, we replicate the findings from Studies 1 and 2 and explicitly test the mediation mechanism hypothesized.

### 4.3 Study 3: Mediation Test Using Two Controlled Experiments

To robustly investigate the mediating role of threat emotions, we conducted two controlled experiments using Prolific, a validated online research platform. Experiment 3a directly tested the mediating role of threat emotions, while Experiment 3b further examined mediation by experimentally manipulating perceptions of threat through different scenarios.

**4.3.1 Experiment 3a: Direct Testing of Main, Mediation, and Moderation Effects**

**Experimental Design**

Experiment 3a employed a 2 (AI innovation: emphasis on AI innovation vs. control) × 2 (Innovation type: process vs. product) between-subjects experimental design.

**Participants**

Participants were recruited through Prolific (https://www.prolific.com), a platform that provides high-quality data from a diverse and pre-screened pool of over 200,000 active participants. Prolific ensures participant reliability and typically yields rapid responses, generally within two hours following the study launch. Our initial sample included 410 participants aged 18 and older.

Following standard data quality procedures, we excluded participants who failed attention checks or submitted incomplete questionnaires. The final sample comprised 404 respondents ($M_{age}$ = 35.89 years, SD = 13.63; 51% female).

**Method and Procedure**

We employed a between-subjects experimental design in which participants were randomly assigned to either an AI innovation condition or a control group, the experimental procedure comprised four distinct parts.

In the first part, participants were presented with general information highlighting significant advancements in AI technology over the past decade, particularly emphasizing machine learning, robotics, and neural networks. Generative AI systems (e.g., ChatGPT), trained on extensive datasets, demonstrated unprecedented capabilities across multiple domains.

**[Materials about AI technology]** *AI has witnessed rapid advancement in the past decade alongside related technologies such as machine learning, robotics, and neural networks. Among* *these, trained on extensive datasets, the emergence of Generative AI (i.e., ChatGPT) has demonstrated unprecedented capabilities across multiple domains.*

In the second part, participants read condition-specific materials regarding corporate AI technology innovation. Participants in the control group proceeded directly without receiving additional details about corporate investments in AI technology.

**[In the AI technology innovation & product group]** *With the rapid development of AI technology (e.g., ChatGPT), the firm has increased its investment in AI technologies. These investments are primarily used to enhance production and operational processes.*

**[In the AI technology innovation & process group]** *With the rapid development of AI technology (e.g., ChatGPT), the firm has increased its investment in AI technologies. These investments are mainly used to enhance firms' new product development. Consumers will have the opportunity to use new products developed by the firms.*

After participants read their assigned materials, we conducted a manipulation check. Specifically, participants rated their perception of the firm's AI technology innovation using a 7-point scale (1 = improving processes, 7 = developing products), responding to the question: "To what extent do you think this firm invests in AI to improve processes or develop products?"

The third part measured participants' negative emotions using four items adapted from Xie et al. (2015) (e.g., "Right now, I feel fear/anger/disgust/disapproval"). Participants also indicated their likelihood of making complaints using four items adapted from Swan and Oliver (1989) (e.g., "I will complain to the service provider about the service quality").

The final part collected participants' demographic data, including gender, age, and AI-related work experience (Ameen et al., 2022a; Ameen et al., 2023; Zhu & Deng, 2021).

**Results**

*Group Equivalence.* Preliminary analyses confirmed that there were no significant differences across experimental groups concerning demographic variables, including gender ($p = .731$), age ($p = .128$), education ($p = .128$), and income ($p = .093$). Thus, comparisons between groups are valid.

*Manipulation Check.* The manipulation check confirmed the effectiveness of our experimental conditions. The AI innovation group participants reported significantly higher perceptions of firm AI investment than the control group ($M_{\text{AI innovation}}$ = 5.84, SD = 0.86; $M_{\text{Control}}$ = 3.10, SD = 1.83; t = 2.71, p = .003).

*Complaints.* A 2 × 2 ANOVA revealed no significant main effect for AI innovation ($F[1, 400]$ = 1.37, p = .242) but showed a significant main effect for innovation type ($F[1, 400]$ = 5.22, p = .023). Crucially, there was a significant interaction between AI innovation and innovation type ($F[2, 400]$ = 4.08, p = .044).

In the product innovation condition, participants expressed significantly more complaints in the AI innovation group ($M_{\text{AI innovation}}$ = 5.77, SD = 1.58) compared to the control group ($M_{\text{Control}}$ = 4.11, SD = 1.55; $F[1, 400]$ = 4.93, p = .012). In the process innovation condition, the difference in complaints between the AI innovation group ($M_{\text{AI innovation}}$ = 4.33, SD = 1.37) and the control group ($M_{\text{Control}}$ = 4.02, SD = 1.47) was marginally significant ($F[1, 400]$ = 11.25, p = .060). These results support $H_{2a}$ and $H_3$, as illustrated in Figure 3a.

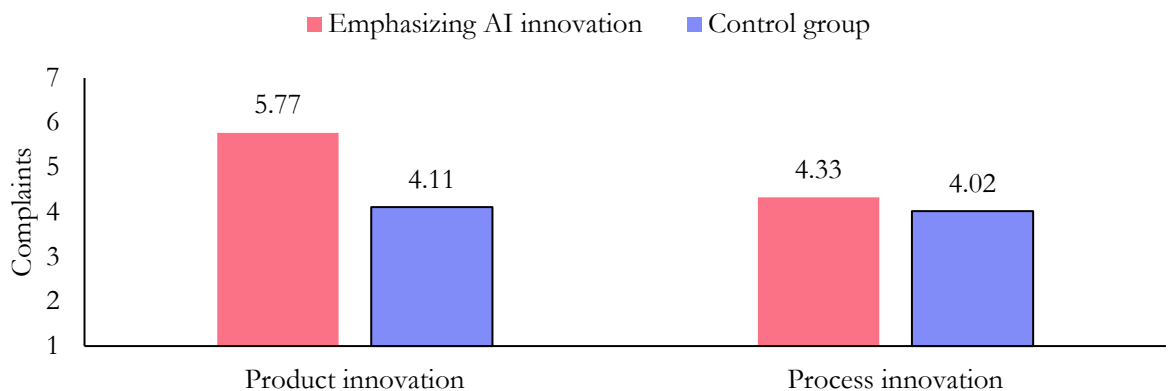

**Figure 3a.** Interaction on Complaints

*The Mediating Effect of Threat Emotion.* We conducted a moderated mediation analysis using PROCESS Model 8 (Hayes, 2013), employing 5,000 bootstrap samples. In this analysis, *AI innovation* served as the independent variable, consumer *complaints* as the dependent variable, threat emotion as the mediator, and *innovation type* as the moderator. As expected, our results indicated that *innovation type* significantly moderated the mediation effect of *threat emotion* on the relationship between *AI innovation* and consumer *complaints* (index = -0.17, SE = 0.11, 95% CI = [-0.4037, -0.0058]). Although the mediating effects of *threat emotion* were significant for both product and process innovations, the effect size was stronger in the product innovation condition compared to the process innovation condition.

These findings support H$_1$, H$_{2a}$, H$_{2b}$ and H$_3$. The final moderated mediation model is illustrated in Figure 3b.

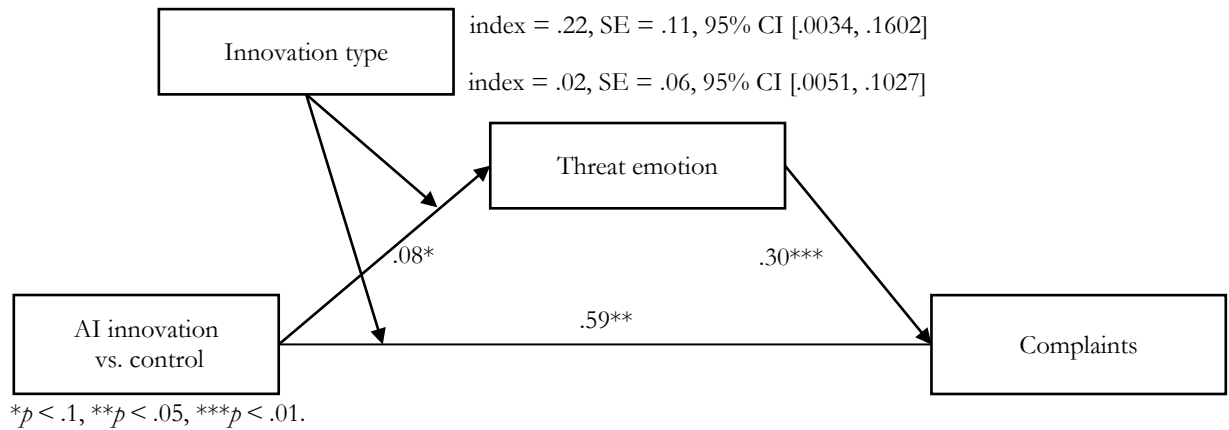

*p < .1, **p < .05, ***p < .01.

**Figure 3b.** The Results of the Mediation Test

### 4.3.2 Experiment 3b: Moderated Mediation Test

**Experimental Design**

Experiment 3b employed a 2 (AI innovation: emphasis on AI innovation vs. control) × 2 (Innovation type: process vs. product) × 2 (Privacy policy: with vs. without) between-subjects experimental design.

**Participants**

We recruited participants through Prolific. The initial sample comprised 500 participants aged 18 and older. Following standard data quality procedures, we excluded participants who failed attention checks or submitted incomplete questionnaires. The final sample comprised 479 respondents (M$_{age}$ = 32.79 years, SD = 10.63, 58% were female).

**Method and Procedure**

Experiment 3b was divided into five distinct parts. The first and final parts were identical to those used in Experiment 3a. However, the fourth section uniquely manipulated participants' perceptions of threat through varying materials. Specifically, participants were provided with company introduction materials designed to vary their perceptions of the firm's commitment to privacy policy. Participants were randomly assigned to either a "with privacy policy" or a "without policy" condition.

**[In the with policy condition]** *Consumer data privacy protection has been a top priority for the firm over the years. It developed a series of policies for consumer privacy protection in response to the growth of AI technology in the past few years.*

**[In the without policy condition]** *Consumer data privacy protection has been a top priority over the years.*

To verify the effectiveness of this manipulation, all participants rated their perception of the firm's commitment to consumer privacy protection using a 7-point scale (1 = strongly disagree, 7 = strongly agree), responding to the question: "To what extent do you think that this firm protects consumer privacy?"

**Results**

*Group Equivalence.* Preliminary analyses confirmed there were no significant differences across experimental groups regarding demographic variables: gender ($p = 0.590$), age ($p = 0.134$), education ($p = 0.158$), and income ($p = 0.257$). Thus, group comparisons were considered valid.

*Manipulation Check.* The manipulation check confirmed the effectiveness of our privacy policy manipulation. Participants in the "with-policy" condition reported significantly higher perceptions of the firm's privacy policy commitment ($M_{\text{with policy}} = 5.29$, $SD = 1.12$) compared to participants in the "without-policy" condition ($M_{\text{without policy}} = 4.29$, $SD = 1.16$; $t = 3.74$, $p = .028$).

*Complaints.* We conducted a 2 (AI innovation: present vs. control) × 2 (Innovation type: product vs. process) × 2 (Privacy policy: with vs. without) ANOVA to examine their effects on consumer complaints. The analysis revealed a significant three-way interaction effect ($F[3, 471] = 5.77$, $p = .031$).

In the group without a privacy policy, participants in the product innovation condition expressed more complaints in the AI innovation condition compared to the control condition. Similarly, those in the process innovation condition displayed a significant difference in the number of complaints between the AI and control conditions. These findings support $H_{2a}$ and $H_3$, as illustrated in Figure 4a.

Conversely, in the group with a privacy policy, participants in the product innovation condition also expressed more complaints in the AI innovation condition compared to the control condition. However, participants in the process innovation condition showed no significant difference in complaints between the AI innovation and control conditions. These differences are illustrated in Figure 4b.

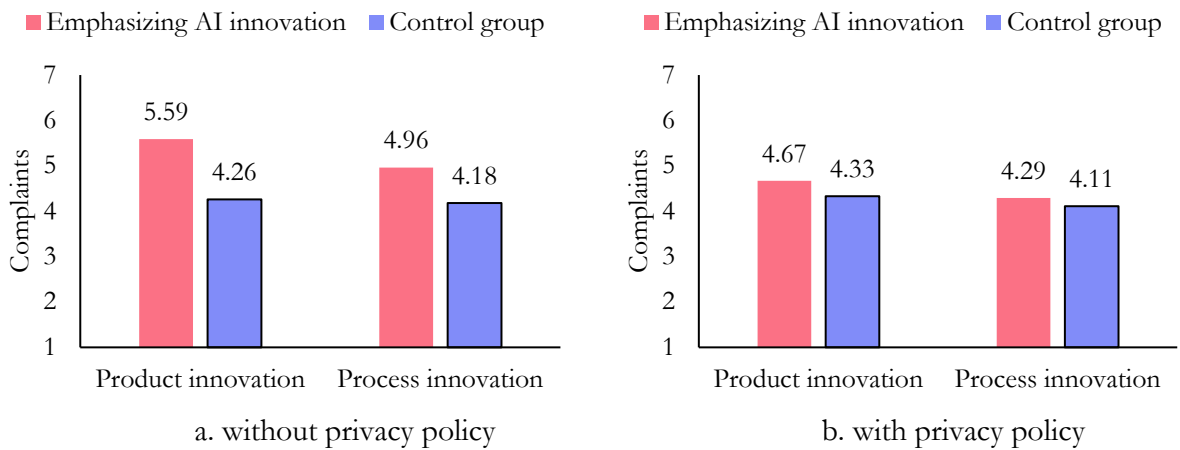

a. without privacy policy      b. with privacy policy
**Figure 4.** Moderated mediation

These results indicate that emphasizing the firm's privacy policy mitigates the impact of AI innovation on consumer complaints, providing further evidence consistent with the mediating role of threat emotions.

## 5. Discussion

In the current age of AI, firms have significantly intensified their investments in AI technology innovation (Trajtenberg, 2018). Given the unprecedented pace of these developments, it is increasingly critical to explore the complex relationship between firms' AI innovation and consumer responses (Chandra et al., 2022; Darveau & Cheikh-Ammar, 2021). Recognizing that consumer complaints can negatively affect brand evaluations and sales (e.g., Babić Rosario et al. 2016), this paper examines the relationship between firms' AI innovation and consumer complaints.

Drawing on Protection Motivation Theory (PMT), we propose and empirically confirm that firms' AI innovation heightens consumers' negative emotional responses to perceived threats posed by AI technology. Moreover, given that product technologies are inherently more visible to consumers than process technologies (Un & Asakawa, 2015), we further propose and find that AI product innovation amplifies consumers' threat emotions, thereby intensifying their complaints.

## 5.1 Theoretical Contributions

This research contributes to the existing literature in three significant ways.

First, we address the emerging challenge of understanding AI's rapid advancement and its dual impact on firm performance (Alderucci et al., 2019; Cockburn et al., 2019; Yang, 2022) and consumer responses (Chandra et al., 2022; Darveau & Cheikh-Ammar, 2021). Our findings reveal how firms' investments in AI technology inadvertently increase consumer complaints. By doing so, we enhance the current understanding of AI's negative consequences (Cheng et al., 2022; Grewal et al., 2021; Grundner & Neuhofer, 2021; Liang et al., 2021), particularly regarding the unprecedented acceleration of AI technology development.

Second, we extend the existing research on different AI innovation types. Few studies have investigated how consumers respond differently to AI innovation empowering intangible processes than tangible products (Frank, 2024; Frank et al., 2021). Drawing on patent classification research (Bena, 2023; Ganglmair et al., 2022), our study explicitly compares AI product and process innovations. We demonstrate that AI product innovation, due to its greater consumer visibility (Un & Asakawa, 2015), increases consumer complaints more significantly than AI process innovation. This finding deepens our theoretical understanding of the nuanced distinctions between product and process innovations (Cohen & Klepper, 1996a, 1996b; Scherer, 1982, 1984).

Third, our findings enhance the understanding of how discrete emotions influence AI and consumer behavior (Blut et al., 2024). This paper highlights limitations in current research on discrete emotions, primarily focusing on broadly categorized positive or negative emotions rather than specific emotional states (Arce-Urriza et al., 2025). Although previous studies have explored discrete emotions such as pleasure (Wang et al., 2007; Clarkson et al., 2013; Marjerison et al., 2022), gratitude (Blut et al., 2024), pride (Fix et al., 2006), anger (Li and Huang, 2020), fear and anxiety (Mori et al., 2012), worry (Ellway, 2016), and contempt (Xia et al., 2004), the specific role of discrete threat-related emotions remains underexplored. We explore the mediating role of discrete threat emotions (e.g., fear, anger, disgust, disapproval) in the relationship between AI innovation and consumer complaints.

Finally, we contribute to the consumer complaint literature within the AI context. As firms increasingly invest in AI technologies and pursue patent developments to sustain competitive advantages (Alderucci et al., 2019; Yang, 2022), the relationship between AI innovation and consumer privacy concerns remains underexplored. Our research illuminates this crucial connection, enriching theoretical insights into consumer complaint behaviors in the rapidly evolving AI landscape.

**5.2 Practical Implications**

This research offers valuable insights for policymakers and business practitioners.

For policymakers, our findings underscore the importance of balanced regulatory approaches to AI technology development to mitigate its potential negative consequences (Martin & Murphy, 2017). Policymakers should create comprehensive regulatory frameworks that foster technological advancement while actively addressing consumers' negative emotional responses. Specifically, our results highlight the need for policies to reduce threat-related consumer emotions, thereby minimizing complaints.

For business practitioners, our findings highlight the crucial balance firms must maintain between investing in AI technologies and managing consumer complaints. Although firms increasingly pursue AI innovation through substantial investments and patent applications to secure competitive advantage (Alderucci et al., 2019; Yang, 2022), they concurrently face significant challenges due to negative consumer responses. Our research indicates that AI product innovation, due to their higher visibility, tends to evoke stronger negative emotions and thus increase consumer complaints more than AI process innovation. Practitioners should thus consider carefully balancing their AI technology investments, perhaps emphasizing transparency and clearly communicating privacy policies, to effectively mitigate consumer threats and associated complaints.

**5.3 Limitations and Directions for Future Research**

Although this study examines the impact of firms' AI technology innovation on consumer complaints using a comprehensive mixed-method approach, several limitations and opportunities for future research remain.

First, while we employed the RoBERTa-base model to classify 28 emotional dimensions (Hartmann et al., 2023; Liu et al., 2019) and confirmed specific negative emotions as predictors of threat perception (Moody et al., 2018), future studies should explore a broader spectrum of emotional responses. Prior research suggests that positive emotions can differentially influence consumer attitudes and behaviors (Lavega et al., 2011). Investigating positive emotions could provide further insights into how discrete emotions influence the relationship between AI technology and consumer responses.

Second, our research focused specifically on consumer complaints as an outcome measure. Future researchers could examine additional downstream behavioral consequences, such as advertising effect (Z. Wang et al., 2023) and consumers' data-sharing practices or purchasing intentions (Li & Xu, 2022). Exploring these outcomes would offer a more comprehensive understanding of how firms' AI technology investments impact consumer behavior beyond complaints alone.

Cohen, W. M., & Klepper, S. (1996b). A Reprise of Size and R&D. *The Economic Journal*, *106*(437), 925-951.

Cram, W. A., D'Arcy, J., & Proudfoot, J. G. (2019). Seeing the Forest and the Trees: A Meta-Analysis of the Antecedents to Information Security Policy Compliance. *Mis Quarterly*, *43*(2), 525-554.

Crossan, M. M., & Apaydin, M. (2010). A Multi-Dimensional Framework of Organizational Innovation: A Systematic Review of the Literature. *Journal of Management Studies*, *47*(6), 1154-1191.

Damanpour, F. (2010). An Integration of Research Findings of Effects of Firm Size and Market Competition on Product and Process Innovations. *British Journal of Management*, *21*(4), 996-1010.

Darveau, J., & Cheikh-Ammar, M. (2021). The interplay between liminality and consumption: A systematic literature review with a future research agenda. *International Journal of Consumer Studies*, *45*(4), 867-888.

Floyd, D. L., Prentice-Dunn, S., & Rogers, R. W. (2000). A Meta-Analysis of Research on Protection Motivation Theory. *Journal of Applied Social Psychology*, *30*(2), 407-429.

Frank, B. (2024). Consumer preferences for artificial intelligence-enhanced products: Differences across consumer segments, product types, and countries. *Technological Forecasting and Social Change*, *209*, 123774.

Frank, B., Herbas-Torrico, B., & Schvaneveldt, S. J. (2021). The AI-extended consumer: Technology, consumer, country differences in the formation of demand for AI-empowered consumer products. *Technological Forecasting and Social Change*, *172*, 121018.

Ganglmair, B., Robinson, W., & Seeligson, M. (2022). The Rise of Process Claims: Evidence From a Century of U.S. Patents. *SSRN Electronic Journal*.

Gans, J. S. (2024). Market power in artificial intelligence.

Gerards, J. (2019). The fundamental rights challenges of algorithms. *Netherlands Quarterly of Human Rights*, *37*(3), 205-209.

Granulo, A., Fuchs, C., & Puntoni, S. (2019). Psychological reactions to human versus robotic job replacement. *Nature Human Behaviour*, *3*(10), 1062-1069.

Gregoire, Y., Khamitov, M., Carrillat, F. A., & Rohani, M. (2024). The attenuation effects of time and "sensemaking" surveys on customer revenge. *Journal of the Academy of Marketing Science*.

Grewal, D., Guha, A., Satornino, C. B., & Schweiger, E. B. (2021). Artificial intelligence: The light and the darkness. *Journal of Business Research*, *136*, 229-236.

Grewal, D., Satornino, C. B., Davenport, T., & Guha, A. (2024). How generative AI Is shaping the future of marketing. *Journal of the Academy of Marketing Science*.

Grundner, L., & Neuhofer, B. (2021). The bright and dark sides of artificial intelligence: a futures perspective on tourist destination experiences. *Journal of Destination Marketing & Management*, *19*, 100511.

Gu, W., Chan, K. W., Kwon, J., Dhaoui, C., & Septianto, F. (2023). Informational vs. Emotional B2B Firm-generated-content on Social Media Engagement: Computerized Visual and Textual Content Analysis. *Industrial Marketing Management*, *112*, 98-112.

Hartmann, J., Heitmann, M., Siebert, C., & Schamp, C. (2023). More than a Feeling: Accuracy and Application of Sentiment Analysis. *International Journal of Research in Marketing*, *40*(1), 75-87.

Hermann, E., & Puntoni, S. (2024). Artificial intelligence and consumer behavior: From predictive to generative AI. *Journal of Business Research*, *180*, 114720.

Hilary, G., & Hui, K. W. (2009). Does Religion Matter in Corporate Decision Making in America? *Journal of financial economics*, *93*(3), 455-473.

Huang, M.-H., & Rust, R. T. (2024). The Caring Machine: Feeling AI for Customer Care. *Journal of Marketing*, 00222429231224748.
31

# Web Appendix

# Understanding the Relationship Between Firms' AI Technology Innovation and Consumer Complaints

## Web Appendix A: Supplementary Materials for RoBERTa

**Emotional Measurement**

Our emotional analysis utilizes the RoBERTa-base model, fine-tuned on the go_emotions dataset, which employs multi-label classification to detect 28 distinct emotional states (Hartmann et al., 2023; Liu et al., 2019). This model, built upon the RoBERTa-base architecture, was specifically fine-tuned for multi-label emotion classification using the AutoModelForSequenceClassification framework, with a learning rate of 2e-5 and weight decay of 0.01 across three training epoch (HuggingFace, 2021).

The go_emotions dataset, constructed from Reddit text data, provides extensive coverage of emotional expressions through its 28-emotion label system (see Table A1). Each input text may correspond to multiple emotional labels, with the model outputting a probability score for each emotion category. Although a standard threshold of 0.5 for probability scores is typically employed, threshold optimization per emotion category has demonstrated improved classification performance. Performance evaluations indicate strong overall classification accuracy, achieving weighted precision of 0.572, recall of 0.677, and an F1 score of 0.611 across all emotional categories.

Table A1. Discrete Emotion Indicators Identified by Roberta and Their Means

| Emotion | Mean (N = 147,060) | Emotion | Mean (N = 147,060) |
|---|---|---|---|
| Emotion_neutral | 0.758232069 | Emotion_anger | 0.000894883 |
| Emotion_admiration | 0.007664306 | Emotion_realization | 0.053249435 |
| Emotion_pride | 0.000518090 | Emotion_confusion | 0.033325555 |
| Emotion_approval | 0.185774407 | Emotion_disgust | 0.001258307 |
| Emotion_optimism | 0.015933182 | Emotion_nervousness | 0.000845695 |
| Emotion_excitement | 0.00171078 | Emotion_grief | 0.000358653 |
| Emotion_gratitude | 0.001423206 | Emotion_fear | 0.001149165 |
| Emotion_curiosity | 0.045662452 | Emotion_joy | 0.001137788 |
| Emotion_relief | 0.001022024 | Emotion_amusement | 0.000706617 |
| Emotion_love | 0.001566607 | Emotion_sadness | 0.002337074 |
| Emotion_surprise | 0.003783527 | Emotion_remorse | 0.000760444 |
| Emotion_caring | 0.002814601 | Emotion_disappointment | 0.009511668 |
| Emotion_embarrassment | 0.000696074 | Emotion_disapproval | 0.014761799 |
| Emotion_desire | 0.002754660 | Emotion_annoyance | 0.007849027 |

# Web Appendix B: Additional Materials for Alternative Measures

**Alternative Measure of IV**

**Patent Classifications of AI-related Innovation**

We employed raw counts of AI-related patents based on International Patent Classification (IPC) codes as an alternative measure of AI technology development (Yang, 2022). Detailed IPC codes are presented in Table B1.

Table B1. Patent Classifications of AI-related Technology Based on the IPC Code

| IPC codes |
|---|
| G05B13/02, G05B13/04, G05B15/0002, G05B19/418, G05B19/042, G06E1/00, G06E3/00, G06F3/01-3/0489, G06F3/143/15, G06F9/44, G06F9/50, G06F9/54, G06F15/18, G06F15/00, G06F17/00, G06F17/20-28, G06F17/30, G06F17/50, G06G7/00, G06J1/00, G06K9, G06K11, G06N3/00, G06N3/02, G06N3/04, G06N3/08, G06N3/10, G06N3/12, G06N5/00, G06N5/02, G06N5/04, G06N7/00, G06N7/02, G06N7/04, G06N7/06, G06N7/08, G06N99/00, G06Q10, G06Q30, G06Q50/04, G06Q90, G06T7, G06T11/80, G06T13, G06T15, G06T17-19, G08G1/0962-0969, G10L13/027, G10L15, G10L17, G10L25/00, G10L25/63, 66, G16H. |

We tested the main effect of AI technology development through two models. Model 1 evaluated the effect independently, while Model 2 included control variables and fixed effects. Model 2 revealed a significant positive coefficient ($\beta = 0.098$, $p = .032$), supporting $H_{2a}$.

Subsequent moderation analyses (Models 3–6) assessed the interaction effects of technology type and privacy policy. Results demonstrated significant moderation effects for both factors, supporting $H_3$ and $H_4$.

**Table B2.** Results of GLM Regression Using Alternative IV

| | Model 1 | Model 2 | Model 3 | Model 4 |
|---|---|---|---|---|
| ROA | | 0.140 | 0.063 | 0.065 |
| | | (0.182) | (0.153) | (0.156) |
| | | [0.439] | [0.682] | [0.677] |
| Size | | -0.019 | 0.136 | 0.140 |
| | | (0.132) | (0.118) | (0.122) |
| | | [0.884] | [0.249] | [0.251] |
| Age | | 0.026 | 0.140 | 0.142 |
| | | (0.161) | (0.158) | (0.161) |
| | | [0.871] | [0.375] | [0.377] |
| Other patents | | 0.211** | 0.217*** | 0.218*** |
| | | (0.094) | (0.077) | (0.079) |
| | | [0.025] | [0.005] | [0.006] |
| Market cap | | 0.836*** | 0.786*** | 0.787*** |
| | | (0.155) | (0.140) | (0.140) |
| | | [0.000] | [0.000] | [0.000] |
| Growth | | -12.851 | -17.425 | -17.798 |
| | | (14.252) | (14.911) | (15.519) |
| | | [0.367] | [0.243] | [0.251] |
| ISO | | 0.434 | 0.288 | 0.285 |
| | | (0.364) | (0.366) | (0.369) |
| | | [0.232] | [0.431] | [0.440] |
| CSR reporting | | 0.054 | -0.068 | -0.081 |
| | | (0.481) | (0.497) | (0.537) |
| | | [0.910] | [0.891] | [0.879] |
| AI innovation _IPC | 0.247*** | 0.098*** | 0.103*** | 0.112** |
| | (0.081) | (0.001) | (0.001) | (0.018) |
| | [0.002] | [0.003] | [0.001] | [0.046] |
| Innovation type | | | -2.174*** | -2.179*** |
| | | | (0.517) | (0.513) |
| | | | [0.000] | [0.000] |
| AI innovation_IPC x Innovation type | | | | 0.079** |
| | | | | (0.435) |
| | | | | [0.046] |
| _cons | -2.212*** | -30.793*** | -30.500*** | -30.549*** |
| | (0.163) | (1.337) | (1.217) | (1.465) |
| | [0.000] | [0.000] | [0.000] | [0.000] |
| Year FE | No | Yes | Yes | Yes |
| Industry FE | No | Yes | Yes | Yes |
| Observations | 4,884 | 2,758 | 2,671 | 2,671 |
| Log likelihood | -2413.991 | -804.699 | -746.840 | -746.798 |

*Notes*: Robust standard errors clustered at the firm-level are reported in parentheses, and exact p-values are reported in square brackets.
Standard errors in parentheses; * $p < 0.1$, ** $p < 0.05$, *** $p < 0.01$.